\documentclass[%
 reprint,
 amsmath,amssymb,
 aps,
floatfix,
]{revtex4-2}

\usepackage{amsmath}

\usepackage{xcolor}
\usepackage{graphicx}% Include figure files
\usepackage{dcolumn}% Align table columns on decimal point
\usepackage{bm}% bold math
\usepackage{amsmath}
\usepackage{mathtools}
\usepackage{enumitem}
\usepackage{xcolor}
\usepackage{graphicx}% Include figure files
\usepackage{dcolumn}% Align table columns on decimal point
\usepackage{bm}% bold math
\usepackage{float}

\begin{document}

\preprint{APS/123-QED}

%\title{
%An Integrated Approach for Evaluating Higher-Order Dependencies Across Different Levels of Resolution: Applications to Cardiovascular and Brain Dynamic Networks
%}

\title{%An Integrated Approach for Evaluating Higher-Order Dependencies Across Different Levels of Resolution
%Investigating Memory Effects in Event-Data: A
%Point-Process Approach with Applications from
%Brain and Autonomic Neuroscience
A Model-Free Method to Quantify Memory Utilization in Neural Point Processes 
}


\author{Gorana Mijatovic}
 \altaffiliation{Faculty of Technical Sciences, University of Novi Sad, Serbia}
 \email{gorana86@uns.ac.rs}

\author{Sebastiano Stramaglia}
 \altaffiliation{Department of Physics, University of Bari Aldo Moro, and INFN Sezione di Bari, Italy}
\email{sebastiano.stramaglia@ba.infn.it }
 
\author{Luca Faes}
\altaffiliation{Department of Engineering, University of Palermo, Italy}
\email{luca.faes@unipa.it}


\begin{abstract}
Quantifying the predictive capacity of a neural system, intended as the capability to store information and actively utilize it for dynamic system evolution, is a key component of neural information processing. Information storage (IS), the main measure quantifying the active utilization of memory in a dynamic system, is only defined for discrete-time processes. While recent theoretical work laid the foundations for the continuous-time analysis of the predictive capacity stored in a process, methods for the effective computation of the related measures, particularly the IS, are needed to favor widespread utilization on neural data. 
This work introduces a method for the model-free estimation of the so-called memory utilization rate (MUR), the continuous-time counterpart of the IS, specifically designed to quantify the predictive capacity stored in neural point processes. The method employs nearest-neighbor entropy estimation applied to the inter-spike intervals measured from point-process realizations for quantifying the extent of memory used by a neural spike train. Moreover, an empirical procedure based on surrogate data is implemented to compensate the estimation bias and to detect statistically significant levels of memory in the analyzed point process.
The method is first validated in simulations of Poisson processes, both memoryless and with memory, as well as in realistic models of coupled cortical dynamics and heartbeat dynamics.
It is then applied to real spike trains reflecting central and autonomic nervous system activities: in spontaneously growing cortical neuron cultures, the MUR detected increasing levels of memory utilization across maturation stages, associated to the emergence of bursting synchronized activity; in the analysis of the neuro-autonomic modulation of human heartbeats, the MUR reflected the sympathetic activation and vagal withdrawal occurring with postural stress but not with mental stress.
The proposed approach offers a novel, computationally reliable tool for the analysis of spike train data in computational neuroscience and physiology.
%model-free and inherently nonlinear approach
%The peculiar advantages of the method are its model-free nature, the fact that it depends only on a few non-critical parameters, and its implementation providing bias compensation and significance assessment.
\end{abstract}

\maketitle

\section{Introduction}

The presence and utilization of memory in neural dynamical activity, typically reflected by the fact that information from the past of a neural process will serve to explain a certain amount of the information contained in its future states, is considered a crucial aspect in computational neuroscience \cite{wibral2015bits,sporns2016networks}. This predictable information is closely tied to the theory of predictive coding, which posits that a neural system continuously generates predictions about incoming sensory input to adapt internal behavior and processing, and that the neural activity underlying these predictions must inherently exhibit a predictable nature, i.e., non-zero memory \cite{friston2006free, grossberg2013adaptive,kilner2007predictive, gomez2014reduced,wibral2014local}. In addition to describing phenomena relevant to the activity of the central nervous system, the investigation of predictive effects in neural processes is also important and widely used to assess the neuro-autonomic control of the cardiovascular system, typically reflected in the analysis of the heart rate and of its variability \cite{robinson1966control,stein1994heart,chen2012unified}.

Memory effects in neural dynamics are typically explored quantifying how much the past states of the analyzed dynamic system are capable of influencing its future evolution. This is often achieved by means of 
measures of time series predictability implemented through a variety of approaches including linear and nonlinear models as well as non-parametric techniques \cite{le1999nonlinear,porta2006complexity,faes2008method,faes2017information,xiong2017entropy}.
Among these approaches, model-free techniques provided in the framework of information theory are highly desirable as they can generalize to virtually any type of dynamics. A well-principled information-theoretic measure quantifying the predictive capability of a dynamic system is the information storage (IS) \cite{lizier2012local,wibral2014local}. Being defined as the information contained in the past history of a stochastic process that is retrieved in its current state, the IS reflects the regularity of the process dynamics resulting from the presence of memory. 
This measure is well-suited to characterize the amplitude of dynamic processes such as those describing the brain wave amplitudes or cardiovascular oscillations.
For instance, IS has been successfully applied to neuronal activity in the central nervous system, including electroencephalographic \cite{faes2016information}, magnetoencephalographic \cite{gomez2014reduced} or other brain signals \cite{wibral2014local}, as well as to cardiovascular and cerebrovascular signals whose amplitude is regulated by the autonomic nervous system \cite{faes2013investigating,javorka2018towards}.

%or based on the concept of the entropy rate, i.e., the conditional entropy of the next outcome given the past state \cite{pincus1991approximate, porta1998measuring}.
%A closely related and complementary measure has gained increasing  attention within the framework of information theory, the amount of information stored in a dynamical system.  

Measures of memory utilization like the IS are well-suited for discrete-time processes, which are usually available in the form of time series data regularly sampled from continuous-time signals or sampled at the rate of a biological clock like the cardiac pacemaker. On the other hand, the application of these measures to point process data represented by sequences of discrete occurrences in continuous time (e.g., neural spike trains or heartbeat events) is not straightforward.
The existing implementations are not well-framed, as they are usually limited to discretize the time axis or to apply the discrete-time formulation to the inter-event series (e.g., interspike or interbeat intervals) treated as continuous-valued signals \cite{chacron2003interspike,faes2013investigating}; as such, they inherently disregard the point-process structure of the underlying events.
Nevertheless, recent theoretical works have defined information-theoretic measures that reflect memory utilization and information exchange in event processes, showing that a rigorous formulation of such measures must be based on computing \textit{information rates} \cite{spinney2017transfer,spinney2018characterizing}. As regards the predictive dynamics within a point process, Spinney et al.~\cite{spinney2017transfer} demonstrated that the rate of IS is a divergent quantity, yet showing that it can be decomposed into two distinct quantities: the instantaneous predictive capacity, which inherits the divergent properties of the IS, and the so-called memory utilization rate (MUR), quantifying the active utilization of memory in the process. Hence, these works identified the MUR as the most appropriate measure for assessing memory effects in point-process data, also underscoring the need for a data-efficient estimator to accurately measure this quantity.

%describing the instantaneous predictive capacity (IPC) and the memory utilization (MU) \cite{spinney2018characterizing} in the process. IPC measures the information storage from the immediately preceding time step (one lag) and reflects the divergent nature of information storage, while MU provides a convergent rate quantity that quantifies the additional accumulation of information storage from time steps prior to the immediate preceding step, offering a more intuitive representation of memory as a rate \cite{spinney2018characterizing}. Since the MU rate is the only component that approaches a limiting value as the discrete time step approaches zero, it is the most suitable measure for investigating memory effects in point processes, i.e., spiking data. 

%All methods treat the R-R interval or heart rate series as continuous-valued signals, rather than model them to reflect the point-process structure of the underlying R-wave events. That is, the R-wave events, which mark the electrical impulses from the heart’s conduction system that represent ventricular contractions,

%thewidemajorityofthesestudiesuseeither beat series (tachograms) unevenly distributed in time,or they interpolate these series with filters notsupportedbyanunderlyingmodelof heartbeat generation.

Building on the rigorous theoretical foundations outlined above, the aim of the present study is to introduce a method for the practical data-efficient quantification of the concept of memory utilization in neural point processes. To reach this goal, we take inspiration from recent methodological advances in the analysis of pairwise interactions between neural spike trains \cite{shorten2021estimating, mijatovic2021information}, embedding them into a computational strategy for the estimation of the MUR in spike process data. Our estimator employs a precise nearest-neighbor entropy estimation method \cite{kraskov2004estimating} applied to inter-spike intervals measured from point process realizations,
%as continuous variables to fully capture the spiking process, 
thereby providing a model-free and inherently nonlinear approach.
Moreover, we introduce an empirical procedure based on surrogate data generation which serves the two-fold purpose of (i) mitigating the estimation bias through the computation of a corrected MUR (cMUR) measure which enforces near-zero values for memoryless processes, and (ii) assessing the significance of the cMUR estimates through statistical testing under the null hypothesis of absence of memory.
The proposed method is first validated in simulations of spike train processes, both memoryless and with memory, as well as in a realistic spiking model of coupled cortical dynamics and in a physiologically-based model of heartbeat dynamics.
Subsequently, it is applied to real spike train data from an animal model of spontaneously growing cortical neuron cultures and from human heartbeat timings assessed in a resting state and during postural and mental stress, in order to investigate the extent of memory involved in different stages of maturation of the central nervous system and in different conditions of activation of the autonomic nervous system. 

The codes implementing the estimation approach introduced in this work are collected in the cMUR Matlab toolbox, which is freely available for download from the repository https://github.com/mijatovicg/cMUR.

\section{Methods}

\subsection{Information Storage in Discrete Time}

Let us consider a dynamical system $\mathcal{X}$ whose temporal evolution can be represented by the discrete-time stochastic process $X = {X_{t_n}}$, where $t_n = n\Delta t$ denotes the time at which the process is sampled; $n \in \mathbb{Z}$ and $\Delta t$ represents the time interval between consecutive samples expressed in unit of time. In the information-theoretic framework, to quantify the amount of information stored in the system $\mathcal{X}$, we use the measure of information storage (IS), expressed as $S_{X} = I(X_{t_i};X^-_{t_i})$, where $X_{t_i}=X_{i\Delta t}$ represents the current state of $X$, $X_{t_i}^-={X_{(i-1)\Delta t},X_{(i-2)\Delta t},\ldots}$ represents its past history, and the symbol $I(\cdot;\cdot)$ denotes the mutual information (MI) between two random variables. The IS quantifies how well the current state of the process can be inferred from its past, thereby reflecting the level of predictability (regularity) of the process dynamics \cite{xiong2017entropy}. For an ergodic and stationary discrete-time process $X$, the IS is defined as \cite{xiong2017entropy}:
\begin{equation}
S_{X} =  \mathbb{E}\left[ \ln \frac{p(x_i|x_i^-)}{p(x_i)} \right] = \lim_{N\to\infty} \frac{1}{N} \sum_{i=1}^{N} \ln \frac{p(x_i|x_i^-)}{p(x_i)},
\label{eq_SE_d}
\end{equation}

\noindent{where $\mathbb{E}[\cdot]$ is the expectation operator, $p(\cdot)$ and $p(\cdot|\cdot)$ denote marginal and conditional probability density respectively, $x_i$ and $x_i^-=\{x_{i-1},x_{i-2},\ldots\}$
are realizations of $X_{t_i}$ and $X_{t_i}^-$, and $N$ is the number of time series points.}

\subsection{Information Storage for Spiking Processes}

The information storage, as defined in  Eq. (\ref{eq_SE_d}), is constrained to analysis of systems operating in discrete time, making it well-suited for processes characterized by time series realizations measured at regularly sampled time points with an interval of $\Delta t$. Since many systems of great interest operate in continuous time ($\Delta t$ tends to zero), some recent theoretical studies have emphasized the importance of normalizing information-theoretic measures to $\Delta t$, so as to obtain the quantification of information rates in [nats/sec] and ensure the convergence of measures as $\Delta t$ approaches zero \cite{spinney2018characterizing, spinney2017transfer}. Nevertheless, normalizing the IS measure when $\Delta t$ tends to zero still produces a divergent rate quantity \cite{spinney2018characterizing}. To address this issue, it is crucial to decompose the IS into two distinct components: one reflecting the active memory usage, known as the memory utilization rate (MUR), and the other representing the instantaneous predictive capacity (IPC). In continuous time, the divergent properties of the IS rate are solely attributed to the IPC term, while the MUR provides a clear and convergent representation of the actual memory utilization in the process \cite{spinney2018characterizing}. In the following section, we will start with the definition of the MUR quantity specifically designed for a specific class of continuous-time processes, namely spiking processes, and then propose a detailed strategy for its data-efficient estimation.

%\begin{equation}
%S_{X} = I_{X} + \dot{M_{X}}\Delta t + O(\Delta t ^2), %\label{eq_SE_dec}
%\end{equation}

%\noindent{where the term $I_{X} \geq 0$ reflects the instantaneous predictive capacity (IPC) of the system, while $\dot{M_{X}} \geq 0$ reflects the effective predictive information and hence it is named the memory utilization rate (MUR). These two quantities are defined as
%\begin{equation}
%I_{X} = \lim_{\Delta t\to 0} \mathbb{E}\left[ \ln %\frac{p(x_i|x_{i-1})}{p(x_i)} \right]},
%\label{eq_I}
%\end{equation}
%\begin{equation}
%\dot{M_{X}} = \lim_{\Delta t\to 0}\frac {0}{\Delta t} \mathbb{E}\left[ \ln \frac{p(x_i|x_i^-)}{p(x_i|x_{i-1})} \right]},
%\label{eq_MUR}
%\end{equation}
%\noindent{where $x_i$,  $x_{i-1}$ and $x_i^-=\{x_{i-1},x_{i-2},\ldots\}$
%are realizations of $X_{t_i}$,$X_{t_{i-1}}$, and $X_{t_i}^-$. So, the $\dot{M_{X}}$ is related to the memory understood in an intuitive manner, while $I_X$ does not characterize the memory, but the predictive capacity that is obtained solely from the current state of the system \cite{spinney2018characterizing}. One should note that consequently the $I_{X}$ does not yield  a rate (it  diverges when $\Delta t \rightarrow 0$) and is not thus a dynamical quantity in the contrast to $\dot{M_{X}}$.}

\subsection{Memory Utilization Rate for Spiking Processes}

%\textcolor {blue} {introduce the conditional instantaneously firing rate}

In the point-process framework, a spiking process is a continuous-time process fully described by  non-overlapping occurrences of indistinguishable events (spikes). Formally, this process is represented by the spike train $X = \{x_{i}\}, \; i=1, 2, \dotsc, N_{X}$, where each $x_{i}$ $\in \mathbb{R}$ denotes the timing of the $i^{th}$ spike, and $N_{X}$ is the total number of spikes in the train. Often, spike trains are studied by examining the sequence of consecutive inter-spike intervals (ISIs): $\{h_{x_i}\}; \; h_{x_{i}} = x_{i} -x_{i-1}, \; i= 2,3, \dotsc, N_{X}$.

The state of a spiking process can be described at each time by the counting process, which counts the number of spikes that have occurred up to that time. For the spike train $X$, the counting process at time $u$ is $N_{X}(u)=n $ if $x_n \leq u <x_{n+1}$. 
The firing rate of a spike train is a measure of the likelihood that the train will produce a spike in a given time interval, relative to the duration of that interval. The instantaneous firing rate for the process $X$ at any time $u$ can be defined by using the counting process as follows:
\begin{equation}
\lambda_{X, u} = \lim_{\Delta{u}\to{0}} \frac{p_u\big(N_X(u+\Delta u) - N_X(u) = 1\big)}{\Delta{u}},
\label{eq_iFR}
\end{equation}
\noindent{where $p_u$ is a probability density evaluated in continuous time (i.e., at any time $u$).}

For the above described spiking process $X$ for which $N_X$ events occur in a period of $T$ seconds, the MUR is defined as \cite{spinney2018characterizing}:
\begin{equation}
\dot{M}_{X} = \lim_{T\to\infty} \frac{1}{T} \sum_{n=1}^{N_X} \ln \frac{\lambda_{X,x_i|h^-_{x_{i}}}}{\lambda_{X,x_i|h_{x_{i}}}} = \overline{\lambda}_{X} \mathbb{E}_{p_{x}}\left[ \ln \frac{\lambda_{X,x_i|h^-_{x_{i}}}}{\lambda_{X,x_i|h_{x_{i}}}}\right],
\label{eq_MUR}
\end{equation}

\noindent{where ${\lambda_{X,x_i|h^-_{x_{i}}}}$ and $\lambda_{X,x_i|h_{x_{i}}}$ are respectively the instantaneous firing rates of the spike train $X$ evaluated at its $i^{th}$ event $x_i$ given the full history ($h^-_{x_{i}} = \{h_{x_{i}}, h_{x_{i-1}}, \cdots\}$) and only the previous event ($h_{x_{i}}$),  and $\overline{\lambda}_{X} = N_{X}/T$ is the average firing rate of the train $X$.} Substituting Eq. (\ref{eq_iFR}) in Eq. (\ref{eq_MUR}) leads to:

\begin{equation}
\small
\dot{M}_{X} = \overline{\lambda}_{X} \\ \lim_{\Delta{u}\to{0}} \mathbb{E}_{p_{x}}\left[ \ln \frac{p_u\big(N_{X}(x_i+\Delta{u}) - N_{X}(x_i) = 1|h_{x_i}^-\big)}{p_u\big(N_{X}(x_i+\Delta{u}) - N_{X}(x_i) = 1 |h_{x_i}\big)}\right].
\label{eq_23}
\end{equation}

\noindent{Then, by applying the Bayes' inversion rule to the conditional probability and recognizing that $\lim_{\Delta{u}\to{0}} p_u\big(\cdot|p_u\big(N_{X}(x_i+\Delta{u}) - N_{X}(x_i) = 1\big)=p_x\big(\cdot\big)$, Eq. (\ref{eq_23}) can be expressed as:
\begin{equation}
\dot{M}_{X} = \overline{\lambda}_{X}\mathbb{E}_{p_{x}} \left[\ln \left(\dfrac{p_{x}(h^-_{x_i})}{p_{u}(h^-_{x_i})} \cdot \dfrac{p_{u}(h_{x_i})}{p_{x}(h_{x_i})}\right)\right],
\label{eq_MUR_hist}
\end{equation}
\noindent{which can be rewritten evidencing four entropy terms as:
 \begin{equation}
\dot{M}_{X} =  \overline{\lambda}_{X} [H_{p_u}(h^-_{x_i})- H(h^-_{x_i}) + H(h_{x_i}) - H_{p_u}(h_{x_i})].
   \label{eq_MUR_H}
\end{equation}

\noindent{It is important to notice that $H(\cdot)$ represents the standard differential entropy, which is computed based on the same probability distribution used for the log-likelihood, $H(h^-_{x_i})=-\mathbb{E}{p_x}[\ln p_x(h^-_{x_i})]$; on the other hand, $H_{p_u}(\cdot)$ is the cross-entropy calculated using two distinct distributions, $H_{p_u}(h^-_{x_i})=-\mathbb{E}_{p_x}[\ln p_u(h^-_{x_i})]$} \cite{shorten2021estimating, mijatovic2021information}.}

\subsection{Estimation of the Memory Utilization Rate}

Eq. (\ref{eq_MUR_H}) makes the basis for estimating the MUR, which relies on the procedure of creating history embeddings through the utilization of ISIs. This procedure, depicted in Fig. \ref{Fig_emb}, distinguishes between the long-term history which includes all events prior to the current spike $x_i$ or any random event $u_i$ (relevant to $H(h_{x_i}^-)$ or $H_{p_u}(h_{x_i}^-)$), and the short-term history which refers to the immediate preceding events from $x_i$ or $u_i$ (relevant to $H(h_{x_i})$ or $H_{p_u}(h_{x_i})$). In practice, when analyzing the long-term history of a spiking event or an arbitrary event, a vector of the last $l$ ISIs is used to approximate such history. For example, if $l=3$, the history at a spiking event $x_i$ is represented by the vector $h^{l=3}_{x_i} = [h_{x_i}, h_{x_{i-1}}, h_{x_{i-2}}]$ (Fig. \ref{Fig_emb}a). At an arbitrary point $u_i$ ($i = 1, 2, ..., N_U$), the long-term history is determined by considering the interval between the most recent preceding spiking event (in the figure $x_{i}$) and $u_i$ ($h_{u_i} = u_i - x_{i}$), along with the previous ($l-1$) ISIs, resulting in the vector $h^{l=3}_{u_i} = [h_{u_i}, h_{x_{i}}, h_{x_{i-1}}]$ (Fig. \ref{Fig_emb}b). The short-term histories represent a special case of the long-term histories when $l=1$, and are given by the vectors $h_{x_i}$ (a) and $h_{u_i}$ (b) in Fig. \ref{Fig_emb}. For a spike train with $N_X$ events $\{x_1, x_2, \cdots, x_{N_X}\}$, the embedding vectors determined as described above form the matrices $\mathbf{H}^l_x$, $\mathbf{H}_x$, $\mathbf{H}^l_u$, and $\mathbf{H}_u$, where $\mathbf{H}^l_x \in \mathbb{R}^{(N'_X \times l)}$ and $\mathbf{H}_x \in \mathbb{R}^{(N'_X \times 1)}$ contain in the $i^{th}$ row respectively the vectors $ h_{x_i}^l$ and $ h_{x_i}$, $ i = l+1, \dots, N_X$, and $\mathbf{H}^l_u \in \mathbb{R}^{(N_U' \times l)}$ and $\mathbf{H}_u \in \mathbb{R}^{(N_U' \times 1)}$ contain in the $i^{th}$ row respectively the vectors $h_{u_{i}}^l$ and $h_{u_{i}}$, $i=l+2, \dots, N_U$. 

 \begin{figure} [h!]
    \centering
    \includegraphics[scale=0.95]{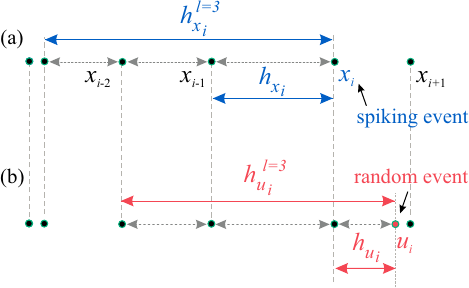}
    \caption{Creation of history embeddings for two types of events. (a) For spiking events ($x_i$), long-history embeddings ($h^l_{x_i}$) are generated using an embedding length of $l$ (in this example $l$ = 3), corresponding to the inter-spike intervals (ISIs) defined by the $l$ spikes preceding $x_i$; short-history embedding ($h_{x_i}$) corresponds to the ISI defined by the single spike immediately preceding $x_i$.  (b) For random events ($u_i$), long-history embeddings ($h^l_{u_i}$) are created using an embedding length of $l$ ($l$ = 3), corresponding to the interval defined by $u_i$ and the preceding spiking event ($x_i$), concatenated with the previous $l-1$ (two in this example) ISIs; the short-history embedding corresponds to the interval defined as $h_{u_i} = u_i - x_i$.}
    \label{Fig_emb}
\end{figure}

These four matrices serve as inputs for the MUR estimation algorithm, which uses the Kozachenko-Leonenko (KL) method \cite{kozachenko1987sample} to estimate the four entropy terms of Eq. (\ref{eq_MUR_H}).  The KL method is based on the $k$-nearest neighbor ($knn$) estimator, an asymptotically unbiased and consistent estimator of the entropy of a multidimensional random variable which utilizes the statistical properties of the distances between its realizations \cite{kozachenko1987sample}. Starting from $N$ realizations of a generic $d$-dimensional variable $W$ forming the data matrix $\mathbf{W}\in \mathbb{R}^{(N \times d)}$, this method estimates the differential entropy $H(W)=-\mathbb{E}_{p_w}[\ln p_w(w)]$ as:
\begin{equation}
\label{HKL}    
\hat{H}(W) = \ln(N-1) -\psi(k) + \ln c_{d} + \dfrac{d}{N} \sum_{i=1}^{N} \ln \varepsilon_{w_i,k,\mathbf{W}},
\end{equation}
\noindent{where $\psi(\cdot)$ is the digamma function, $c_d$ is the volume of the $d$-dimensional unit ball under a given norm, and $\epsilon_{w_i,k,\mathbf{W}}$ is twice the distance between the $i^{th}$ realization of $W$ and its $k^{th}$ nearest neighbor taken from $\mathbf{W}$. To estimate a cross-entropy term $H_{p_u}(W)=-\mathbb{E}_{p_w}[\ln p_u(w)]$ from two data matrices $\mathbf{W}\in \mathbb{R}^{(N \times d)}$ and $\mathbf{V}\in \mathbb{R}^{(M \times d)}$, Eq. (\ref{HKL}) is modified as:}
\begin{equation}
\label{HKLcross}    
\hat{H}_{p_u}(W) = \ln(M) -\psi(k) + \ln c_{d} + \dfrac{d}{N} \sum_{i=1}^{N} \ln \varepsilon_{w_i,k,\mathbf{V}},
\end{equation}
\noindent{where $\epsilon_{w_i,k,\mathbf{V}}$ is twice the distance between the vector $w_i \in \mathbf{W}$ and its $k^{th}$ nearest neighbor taken from $\mathbf{V}$.} 
 
In our case, Eq. (\ref{eq_MUR_H}) involves the estimation of two entropy terms $-$ $H(h_{x_i}^l)$ and $H(h_{x_i})$ $-$ by using Eq. (\ref{HKL}) with $w_i$ corresponding respectively to $h^l_{x_i} \in \textbf{H}^l_{x_i} $ and $h_{x_i} \in \textbf{H}_{x_i} $, 
as well as two cross-entropy terms $-$ $H_{p_u}(h_{x_i}^l)$ and $H_{p_u}(h_{x_i})$ $-$ by using Eq. (\ref{HKLcross}) with $w_i$ corresponding to $h^l_{u_i} \in \textbf{H}^l_{u_i}$ and $h_{u_i} \in \textbf{H}_{u_i}$ (considering that $h_{x_i}^- \approx h_{x_i}^l$ and $h_{u_i}^- \approx h_{u_i}^l$).

Importantly, instead of using a naive estimator that fixes the parameter $k$ and applies Eqs. (\ref{HKL}) and (\ref{HKLcross}) to each of the four entropy terms, we implement the bias compensation strategy as proposed in \cite{shorten2021estimating}. By adjusting the number of neighbors $k$ for each data sample, this strategy enables the use of the same range of distances across spaces of different dimensions, thereby minimizing the bias of the estimates of the sum of entropies in Eq. (\ref{eq_MUR_H}). The algorithm begins by setting a fixed parameter $k_{global}$ which represents the minimum number of nearest neighbors to be used in any search space. For each iteration, with $w_i$ being a realization from the data matrix $\textbf{W}$, the algorithm performs both neighbor and range searches. In the neighbor search, $k$ is fixed while the distance $\varepsilon_{w_i,k,\mathbf{W}}$ is calculated between $w_i$ and its $k^{th}$ nearest neighbor within $\textbf{W}$, or within a  data matrix $\textbf{V}$ in the case of cross-entropy estimation.  In the range search, the number of neighbors $k_{w_i,\mathbf{W}}$ of $w_i$ found within $\textbf{W}$ (or $\textbf{V}$) is counted.
By summing up the four terms obtained utilizing Eq. (\ref{HKL}) twice for the entropy terms and Eq. (\ref{HKLcross}) twice for the cross-entropy terms, we derive the final expression for the MUR estimator:
%\begin{equation}
%\hat{\dot{M}}_{X}= \dfrac{\overline{\lambda}_X}{N'_X} \sum_{i=1}^{N'_{X}}   \psi(k_{h_{x_i},\mathbf{H}_u}) -\psi(k_{h_{x_i},\mathbf{H}_x}) + \psi( k_{h^l_{x_i}, \mathbf{H}^l_x})-\psi(k_{h^l_{x_i},\mathbf{H}^l_u}) \\
 %+  l \ln \dfrac{\epsilon_{h_{x_i},k_{h_{x_i},\mathbf{H}_x},\mathbf{H}_x} \cdot \epsilon^2_{h^l_{x_i},k_{h^l_{x_i},\mathbf{H}^l_u},\mathbf{H}^l_u}}{\epsilon_{h_{x_i},k_{h_{x_i},\mathbf{H}_u},\mathbf{H}_u} \cdot \epsilon^2_{h^l_{x_i},k_{h^l_{x_i},\mathbf{H}^l_x},\mathbf{H}^l_x}}. 
 %\label{final_MUR_est}
%\end{equation}
%\begin{strip}
\begin{equation}
\begin{aligned}
\hat{\dot{M}}_{X}= \dfrac{\overline{\lambda}_X}{N'_X} & \sum_{i=1}^{N'_{X}}   \psi(k_{h_{x_i},\mathbf{H}_u}) - \psi(k_{h_{x_i},\mathbf{H}_x}) \\
& + \psi( k_{h^l_{x_i}, \mathbf{H}^l_x}) - \psi(k_{h^l_{x_i},\mathbf{H}^l_u}) \\
& + l \ln \dfrac{\epsilon_{h_{x_i},k_{h_{x_i},\mathbf{H}_x},\mathbf{H}_x} \cdot \epsilon^2_{h^l_{x_i},k_{h^l_{x_i},\mathbf{H}^l_u},\mathbf{H}^l_u}}{\epsilon_{h_{x_i},k_{h_{x_i},\mathbf{H}_u},\mathbf{H}_u} \cdot \epsilon^2_{h^l_{x_i},k_{h^l_{x_i},\mathbf{H}^l_x},\mathbf{H}^l_x}}.
\end{aligned}
\label{final_MUR_est}
\end{equation}
%\end{strip}
This model-free estimator of the amount of memory utilization in an event process depends only on the parameter $k_{global}$, determining the number of neighbors of the current point which are looked for in the search for neighbors, and on the parameter $l$ setting the length of the long-history embeddings. The dependence of the estimates on these parameters is investigated in simulated settings in Sect. \ref{sec:validation}.

\subsection{Assessment of statistical significance and bias-compensation.} Accurately estimating information-theoretic measures from finite-length process realizations poses a practical challenge, as these estimates can be subject to bias and variance influenced by various factors, including system dynamics, analysis parameters, and the length of the data being analyzed. In order to identify non-negligible memory effects in a spike train, we assess the statistical significance of the MUR estimates by using a technique based on surrogate data \cite{faes2019comparison}. Specifically, 100 surrogate spike trains were generated for each original train by randomly shuffling its ISIs; this approach preserves the ISI distribution of the original spiking process, while eliminating correlation between its ISIs. The MUR estimate was deemed statistically significant if its value obtained on the original spike train exceeded the $95^\textrm{th}$ percentile of its distribution evaluated on the surrogate spike trains. Moreover, the same surrogates were used also to compensate for the potential bias in the MUR estimate through an approach resembling that presented in \cite{mijatovic2022measuring}. Specifically, a corrected MUR (cMUR) is defined as:
\begin{equation}
c\hat{\dot{M}}_{X} = \hat{\dot{M}}_{X} - \langle{\hat{\dot{m}}^S_{X}}\rangle,
\label{eq_cMUR_est}
\end{equation}
\noindent{where $\hat{\dot{M}}_{X}$ is the MUR estimate obtained from the original spike train $X$, while $\langle{\hat{\dot{m}}^S_{X}}\rangle$ is the median of the MUR estimates obtained from several surrogate trains.  Importantly, utilizing Eq. (\ref{eq_cMUR_est}) ensures compensation for the bias in the case of memoryless processes, i.e., those with uncorrelated ISIs; however, a reasonable correction for the bias can still be achieved even when the samples of the observed process are correlated \cite{papana2011reducing}.

\section{Validation on simulated neuronal and cardiac spiking processes}
\label{sec:validation}

In this section we explore the performances of the MUR estimator when applied to different simulations of spiking dynamics, encompassing both realistic neuronal and cardiac dynamics, studied across different conditions of ISI-dependency. The first simulation, emulating the neuronal activity, produced dynamic patterns featuring both independent and history-dependent ISIs generated from an exponential probability density distribution, thereby spanning from memoryless spiking dynamics to dynamics with non-negligible memory capacity. In each of the 100 iterations, we simulated processes with $N_X$ observations, where $N_X \in \{250, 500, 1000\}$ for mutually independent ISIs and $N_X = 1000$ for processes with mutually dependent ISIs. The second simulation focused on a more biologically-relevant neuronal network scenario by employing the Izhikevich model  to simulate the dynamics of coupled cortical neurons \cite{izhikevich2003simple}. In the third simulation, we replicated typical short-term cardiovascular (setting $N_X$ = 300 beats) by modeling the cardiac inter-beat intervals (where each beat corresponds to a spike) according to the realistic history-dependent inverse Gaussian (HDIG) model proposed in \cite{barbieri2005point}. In all simulations, the MUR estimator was implemented by using a set of  $N_U = N_X$ arbitrary points, uniformly distributed within the interval [0, $T$], while exploring the set of parameters $l \in \{2, 3, 4\}$ and  $k_{global} \in \{5, 15, 25\}$. 

\subsection{Simulation 1}

First, we generated spike trains of length $N_X$ encompassing ISIs that exhibit either mutual independence or various degrees of mutual dependence. This was achieved by generating ISI sequences such that the first ISI was drawn from the exponential distribution, $h_{x_1}\in$ EXP(1/$\lambda$); $\lambda$ = 1 spike/s; then, each successive $\text{ISI}$ was generated with a mean depending on the preceding ISI, $h_{x_i} \in$ EXP($1/\lambda(1-p) + p\cdot h_{x_{i-1}})$. Through a gradual increase of the coefficient $p$ from 0 to 0.9 in steps of 0.1, we fine-tuned the generation of ISI sequences, moving from independent ISIs descriptive of memoryless spiking processes ($\dot{M}_X = 0$) to sequences where ISIs depend progressively more and more on their own past history, descriptive of gradually increasing memory effects which are expected to result in increasing values of $\dot{M}_X$. 

Fig. \ref{fig2}a presents the application of the MUR estimator on spike trains with independent ISIs ($p = 0$) as a function of the parameters $N_X, l$ and $k_{global}$. The results indicate the expected absence of memory, with values of $\hat{\dot{M}}_{X}$ distributed around zero when $k_{global}$ = [15, 25] and $N_X$ = [500, 1000], regardless of the  embedding length $l$. On the contrary, for $k_{global}$ = 5, and particularly for data size $N_X$ = 100, the results exhibit negative bias and higher variance, both more pronounced with an increase in the embedding length. Note that the bias can be reliably addressed by utilizing cMUR; here we computed the non-compensated MUR estimator to display the extent of such bias.
% aimed at providing insight into its magnitude, can be addressed by utilizing cMUR; this approach will be implemented in the third simulation as the bias was not observed in the second simulation.
Nevertheless, the MUR estimator consistently yielded a rate of false-positive presence of memory that remained around the nominal 5 \% significance level, as evidenced by the numbers displayed close to the MUR values in Fig. \ref{fig2}a.

\begin{figure*} [t]
    \centering
    \includegraphics[scale=0.95]{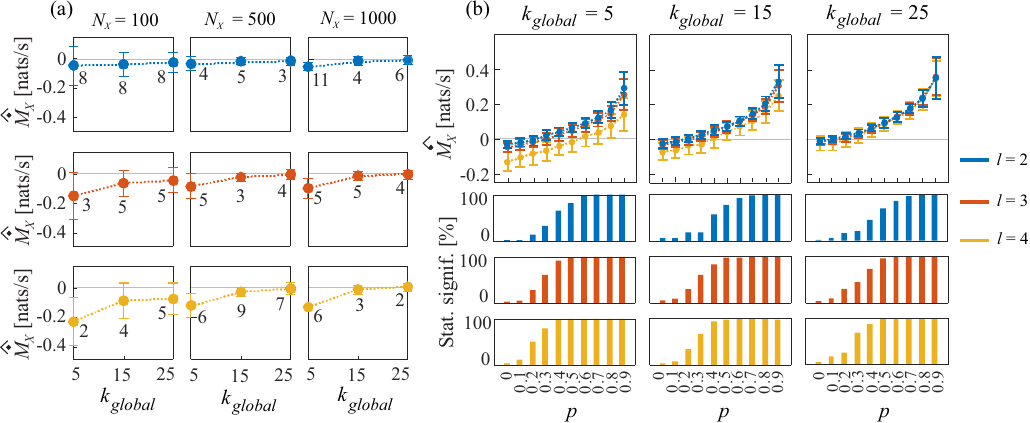}
    \caption{Simulation of spiking dynamics with  independent (a) and history-dependent (b) inter-spike intervals (ISIs) taken from an exponential distribution with parameter $\lambda$ = 1 spikes/s. Plots depict the distribution (mean $\pm$ standard deviation) of the MUR estimates computed over 100 realizations of spiking processes with varying length $N_X = \{100, 500, 100\}$ when ISIs are independent (a), and with a fixed length $N_X = 1000$ when ISIs are history dependent (b), in both cases exploring the parameters $l = \{2, 3, 4\}$ and $k_{global} = \{5, 15, 25\}$. The numbers close to the MUR values in (a) and the bars in (b) report the percentage of realizations for which the MUR estimates were detected as statistically significant using surrogate data.}
    \label{fig2}
\end{figure*}

Fig. \ref{fig2}b displays the MUR estimates as the coupling parameter $p$ moves from 0 to 0.9, and when $N_X$ is fixed to 1000 samples; across all combinations of $k_{global}$ and $l$, the estimates consistently exhibit a progressive increase at increasing the parameter $p$, which mirrors the extent of memory preserved within the process, together with a shift from non-significant to highly significant $\hat{\dot{M}}_{X}$ values. A subtle negative bias emerges, but these estimates lack the statistical significance. 
 
Overall, the results of this simulation highlight the ability of the MUR estimator to effectively capture both  the lack of self-predictable patterns in simulated spike trains with uncorrelated ISIs (Fig. \ref{fig2}a), as well as the presence of self-predictable dynamics and its intensity in spike trains characterized by history-dependent ISIs (Fig. \ref{fig2}b).

\subsection{Simulation 2}

In the second simulation, we used a realistic spiking model to generate coupled-cortical dynamics as proposed in \cite{izhikevich2003simple}. This simulation incorporates two widely recognized models for initiation and spreading of action potentials in neurons, i.e. the Hodgkin-Huxley model and the quadratic integrate-and-fire model.
By receiving both synaptic and noisy thalamic inputs, the simulated dynamics successfully emulate the spiking activity of various fundamental types of neurons found in the mammalian cortex, including regular spiking (RS) excitatory cells and low-threshold spiking (LTS) inhibitory cells \cite{izhikevich2003simple, izhikevich2007dynamical}. Here, we examine a large-scale neural network consisting of 1000 randomly connected cells (800 RS and 200 LTS cells), whereby we evaluate the performance of the MUR estimator by observing that an increased neuronal coupling$-$ reflecting larger synchronization among interconnected neurons$-$ results in a more predictable (regular) dynamics at the level of individual cells.

\begin{figure} []
\centering
    \includegraphics[scale=0.9]{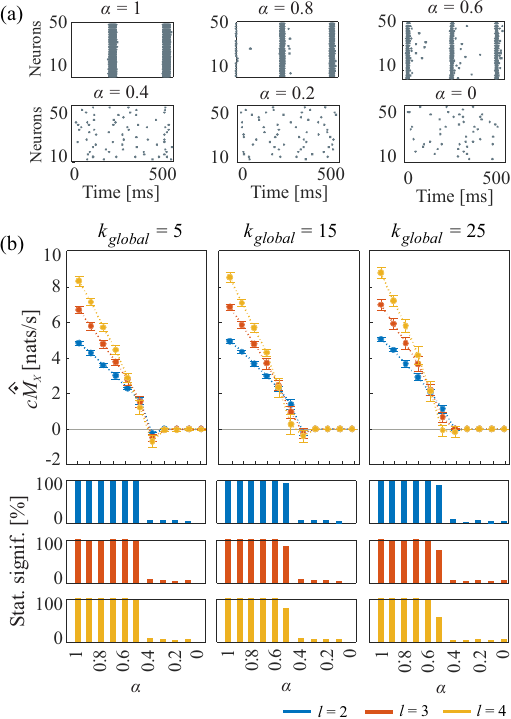}
    \caption{Simulation of RS-LTS neural-network responses with different degrees of synchronicity analyzed computing the MUR with the parameters  $l = \{2, 3, 4\}$ and $k_{global} = \{5, 15, 25\}$. (a) Time-frames of spiking activity for fifty randomly selected RS-LTS neural responses  when network was stimulated with different synaptic inputs modulated by the parameter $\alpha = \{1, 0.8, 0.6, 0.4, 0.2, 0\}$; (b) cMUR estimates expressed as mean $\pm$ standard deviation for the selected responses, and the bar plots representing the statistical significance of the estimates as  $\alpha$ varies.}
    \label{fig3}
\end{figure}

The strength of the synaptic connection between the output of the $j^{th}$ neuron and the input to the $i^{th}$ neuron was modeled by the element $s_{ij} \in $ [-1, 1] of a synaptic matrix $S$. 
The matrix $S$ was initially generated to reflect the maximal synaptic strength among randomly connected neurons, after which the network was simulated several times, each time employing a scaled synaptic matrix $S_{\alpha} = \alpha \cdot S$; $ 0 \geq \alpha \leq 1$, such that initially-established synapses are preserved  but with uniformly reduced strength. Therefore, the parameter $\alpha$ serves to modulate the extent of memory in individual neural responses, since as we gradually decrease this parameter from 1 to 0 we simulate a shift from highly-coupled dynamics with memory ($\dot{M}_X > 0$) to random memoryless dynamics ($\dot{M}_X = 0$).

Due to computational constraints and without sacrificing generality thanks to the homogeneity of the network, we chose to analyze a representative subset of fifty cells randomly selected from the RS-LTS network. The raster plots in Fig. \ref{fig3}a illustrate 
these responses, while Fig. \ref{fig3}b presents the distribution of their MUR estimates together with the bars indicating the proportion of statistically significant estimates determined through the surrogate data analysis.  When neuronal coupling is at its maximum ($\alpha$ = 1), high synchronicity among the neuronal responses is evident (Fig. \ref{fig3}a); this leads to highly regular (self-predictable) individual dynamics which are quantified by the statistically significant $c\hat{\dot{M}}_{X}$ of $\approx$ 9 nats/s (remarkably, this behavior remains consistent across various combinations of $k_{global}$ and $l$, Fig. \ref{fig3}b). As the neuronal coupling gradually decreases, synchronicity among neurons also gradually diminishes (Fig. \ref{fig3}a), leading to a reduced predictability in individual cells; this is captured by statistically significant cMUR estimates between $\approx$ 7 nats/s and 1 nats/s for $\alpha$ respectively between 0.9 and 0.5. When $\alpha$ drops below 0.5, the synaptic links between neurons degenerate yielding to random neural responses lacking of regular firing patterns (Fig. \ref{fig3}a); this shift is effectively detected by the MUR estimates which sharply converged towards statistically non-significant zero values (Fig. \ref{fig3}b).

\subsection{Simulation 3}

In the third simulation, heartbeat occurrence times are generated by following the realistic HDIG model \cite{barbieri2005point}, in a way such that the inter-beat intervals belong to an inverse Gaussian IG ($\mu, \lambda$) distribution, where the mean $\mu$ is dependent on the history of the intervals up to the current heartbeat time according to a linear auto-regressive (AR) model, and the shape parameter $ \lambda$ is fixed (here set to $\lambda = 600$ s). In order to reproduce typical heartbeat interval oscillations within the very low frequency (VLF, $<0.04$ Hz), low frequency (LF, $0.04-0.15$ Hz) and high frequency (HF, $0.15-0.4$ Hz) bands, we use an AR model of order five, whose parameters are set placing two complex-conjugate poles with modulus $\rho_{\text{LF}}$ and phases $ \pm 2\pi\cdot0.1$ rad, two other complex-conjugate poles with modulus $\rho_{\text{HF}}$ and phases $ \pm 2\pi\cdot 0.25$ rad, and a real pole with modulus $\rho_{\text{VLF}} = 0.6$; see \cite{beda2017estimation, stein1994heart,faes2014lag} for details about this type of simulation. 

The simulation was implemented by gradually increasing the parameter $\rho_{\text{LF}}$ from 0.5 to 0.9 in steps of 0.1  while simultaneously decreasing $\rho_{\text{HF}}$ from 0.5 to 0.1, so as to shift from the condition of coexistence of moderate LF and HF oscillations to the condition of amplified LF and attenuated HF in the simulated heartbeat intervals. This setting allowed us to control the modulation of history dependence among intervals and thereby to fine-tune the degree of memory embedded in such dynamics. Given the simulation of a typical 300-beats realizations, in Fig. \ref{sim_cardio} we present the results obtained by applying the cMUR estimator (Eq. \ref{eq_cMUR_est}) as a function of $\rho_{\text{LF}}$.
So, as this parameter increases while $\rho_{\text{HF}}$ simultaneously decreases, leading to a more prominent 
LF-oscillation patterns while HF are damped, the $c\hat{\dot{M}}_{X}$ values also gradually increase, which confirms the ability of the estimator to correctly capture the expected increased predictive capacity within the simulated cardiovascular dynamics. This finding remains consistent across all combinations of $k_{\text{\it global}}$ and $l$ parameters, with deviations observed only for higher values of $\rho_{\text{LF}}$ when $k_{\text{\it global}}=5$ and $l=2$ (see Fig. \ref{eq_cMUR_est}a). Moreover, even though all estimates demonstrate 100 $\%$ statistical significance (results not depicted in the figure), the observed high variance can be ascribed to the small data size \cite{mijatovic2022measuring}.

\begin{figure} [h!]
\centering
    \includegraphics[scale=0.87]{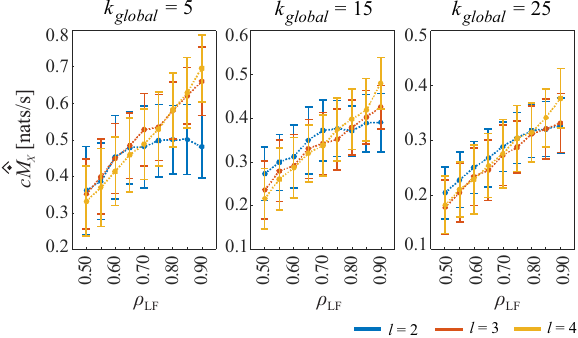}
    \caption{Simulations of realistic heartbeat dynamics based on a history-dependent inverse Gaussian model, analyzed with the cMUR estimator computed with parameters $l = \{2, 3, 4\}$ and $k_{\text{global}} = \{5, 15, 25\}$, while $N_U = N_X = 300$. The statistically significant cMUR estimates show an increasing trend as the parameter $\rho_{\text{LF}}$ increases while $\rho_{\text{HF}}$ concurrently decreases, which accurately identifies a stronger predictive capacity of the simulated heartbeat dynamics.}
    \label{sim_cardio}
\end{figure}

\section{Application to real data}

This section reports the application of the proposed continuous-time MUR estimator to different datasets of experimental recordings, performed with the aim of investigating the presence and extent of memory in neural spike trains related to central activity of the central and autonomic nervous systems.
The first dataset was obtained from a public repository collecting in-vitro recordings of animal cultures of dissociated cortical neurons at various stages of neuronal development \cite{wagenaar2006extremely}. The second dataset includes the heartbeat timing events measured from  healthy subjects at rest and during postural and mental stress \cite{faes2017information}.
Despite their different origins, the utilization of these datasets allowed us to investigate memory capacity in the dynamics of diverse neuro-physiological processes studied at different scales: (i) at a microscopic scale in the central nervous system by analyzing spiking activity within neural cultures as they mature; and (ii) at a peripheral level by exploring the variations in heartbeat timings known to be regulated by the autonomic nervous system.
%To address a potential bias in estimates, mainly arising from the limited duration of the analyzed processes, we applied the cMUR estimator in both applications.

\subsection {Neuronal Spike Trains from In-vitro Neural Cultures}
The dataset utilized in the first application originates from a rich collection of cortical-cell cultures that were extracted from rat embryos and cultured on glass wells under different conditions including both day and night recordings \cite{wagenaar2006extremely}. Our focus was on daily spontaneous recordings obtained from a multi-electrode array (MEA) with an $8 \times 8$ grid of electrodes, at a density of 2500 cells per micro-liter and approximately fifty thousand cells plated on it. Each electrode recorded the spiking activity of an ensemble of roughly one hundred to one thousand neurons, resulting in multi-unit activity (MUA) \cite{wagenaar2006extremely}. A total of 59 MUAs were obtained, with electrodes positioned at a pitch of 200 $\mu$m, except for the ground electrode and those not placed on the corners of the MEA. More details about the protocol can be found in \cite{wagenaar2006extremely}.

In-vitro neuronal cultures are recognized for their ability to self-organize and exhibit synchronous bursting activity that becomes more prominent as they mature \cite{wagenaar2006extremely, pasquale2008self, schroeter2015emergence, chiappalone2006dissociated}. Accordingly, we tested the hypothesis that the emergence of synchronous bursts of neural spiking activity is associated with the rise of self-predictable individual dynamics, and thus increased memory capacity, during neuronal maturation.
To this end, we analyzed MUAs in sixteen cultures by applying the cMUR estimator not only to the overall spiking activity of each MUA, but also at the scale of cellular bursting activity. Burst detection was carried out empirically by using the approach described in \cite{mijatovic2021information}, whereby the spikes within each identified burst were substituted with their respective center of mass, resulting in a final sequence of burst events for each MUA upon which the cMUR estimator was applied.
The cultures were analyzed longitudinally across three stages of maturation, labeled as early ($\approx$ 7 days in vitro, DIV), intermediate ($\approx$ 15 DIV), and mature ($\approx$ 25 DIV).

First, the overall spiking activity was analyzed computing the cMUR over the original sequences of spike trains recorded by each analyzed electrode.
Fig. \ref{culs_boxplot} displays the distribution of the cMUR estimates for the MUA of all sixteen cultures at different stages of culture development. The figure includes both significant individual $c\hat{\dot{M}}_{X}$ values (gray dots) and non-significant ones (red dots), along with the results of the Wilcoxon signed-rank test used to assess the statistical significance of the cMUR differences between pairs of stages. The analysis revealed a consistent trend of increasing cMUR values across stages of development, which was observed for nearly all cultures and was supported by the statistical significance of the cMUR estimates. This increase typically followed a gradual progression from early to intermediate and mature stages, as observed in cultures 1-2, 1-4, 1-5, 2-5, 7-1, 8-2 in Fig. \ref{culs_boxplot}. However, specific cultures, such as 3-1, 3-2, 5-2, and 8-3, exhibit a noticeable increase in $c\hat{\dot{M}}_{X}$ values solely during the transition from the early to the intermediate stage, with no significant alteration observed when switching from the intermediate to the mature stage.

\begin{figure*} [t]
\centering
    \includegraphics[scale=0.95]{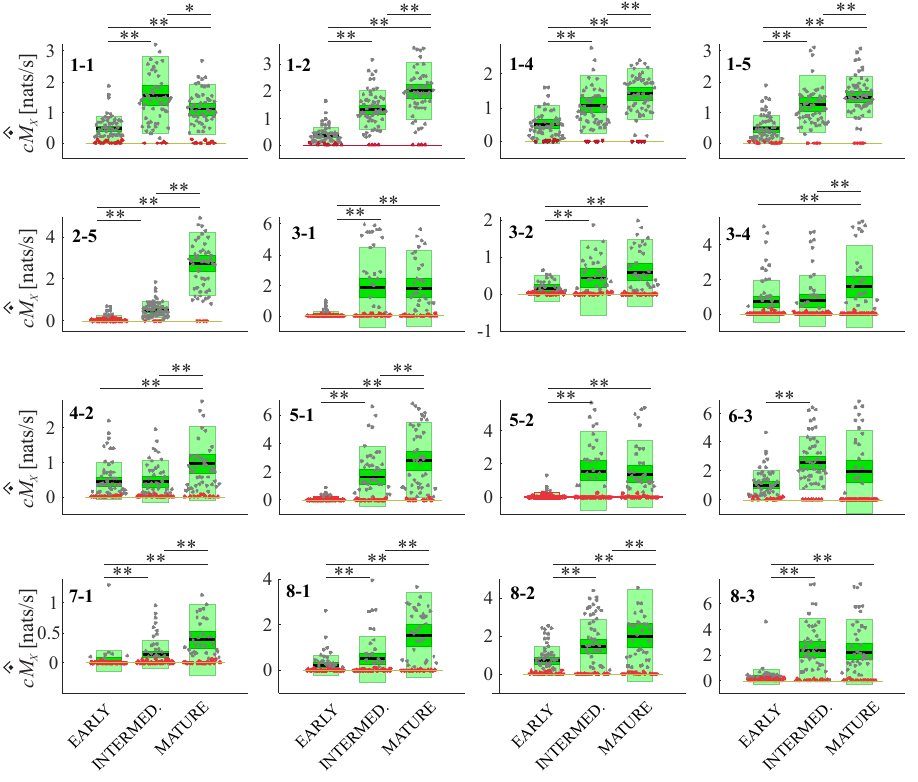}
    \caption{cMUR estimated for the overall spiking activity of sixteen cultures observed at early, intermediate, and mature stages of neural development. Each panel depicts the boxplot distributions and individual values (dots, one for each electrode) of the cMUR estimates; non-significant values detected by surrogate data analysis are denoted by red dots, whereas significant values are shown in gray. The Wilcoxon signed-rank test was used to assess the statistical significance of the differences in the median value of the cMUR estimates between stages, assuming a $5\%$ significance level: $** p<0.005$ and $*p<0.05$. The cMUR estimator was implemented by setting: $k_{global}$ = 25, $l$ = 3, $N_U = N_X$.}
    \label{culs_boxplot}
\end{figure*}

\begin{figure} [!t]
\centering
    \includegraphics[scale=0.97]{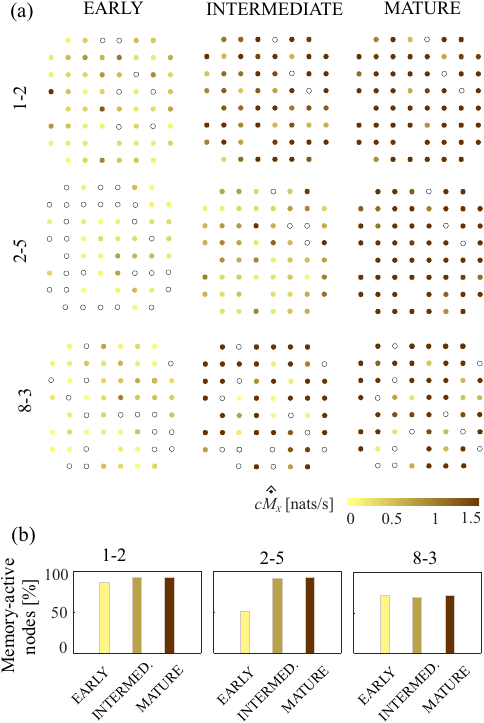}
    \caption{a) cMUR networks constructed for three representative cultures studied at the early, intermediate and mature stages of neuronal development.
     In these networks, color-coded cMUR values are reported at each electrode location exploiting the spatial position of electrodes provided in \cite{data_link}. The electrodes with non-significant cMUR values, as determined by surrogate-data testing, are denoted by empty circles. $c\hat{\dot{M}}_{X}$ values higher than 1.5 nats/s are not shown due to better visualization. The cMUR estimator was implemented by setting the following parameters: $k_{global}$ = 25, $l$ = 3, $N_U = N_X$. b) The percentage of memory-active electrodes (i.e., with a significant $c\hat{\dot{M}}_{X}$ values).}
    \label{culs_nodes}
\end{figure}

Additionally, to investigate how the spatial distribution of the memory capacity of the neuronal cultures evolves as they mature, we utilized a network-graph representation based on the spatial positions of electrodes from the data source \cite{data_link} to depict the cMUR values computed at the level of each electrode.
As shown in Fig. \ref{culs_nodes}a, the cMUR networks for three selected cultures reveal a progressive rise of the magnitude of the cMUR estimates as these cultures move from the early stage of neural development to the intermediate and mature stages. Moreover, defining memory-active electrodes (MAEs) as those whose MUA exhibits statistically significant cMUR values, a substantial proportion of cultures ($\approx$ 68\%) displays the majority of electrodes being active from  the first (early) to the last (mature) stage of neural development.  This is evident in cultures 1-2 and 8-3 in Fig. \ref{culs_nodes}b, where memory-inactive electrodes in the mature stage were inactive also in the  early stage, as evidenced in Fig. \ref{culs_nodes}a. On the other hand, around 22\% of the remaining cultures exhibit an increasing trend in the proportion of MAEs as they mature, or they  maintain a similarly increased number of MAEs throughout the intermediate and mature stages compared to the early stage; an example of this behavior can be observed in culture 2-5 with an electrode layout given in Fig. \ref{culs_nodes}b.

While the previous results refer to the overall spiking activity recorded by the MEA, a markedly different behavior was observed when the analysis was performed on the sequences of burst events. The number of burst events detected in such analysis was considerably smaller than the number of spikes (mean numbers of events per stage observed across all electrodes and cultures [mean $\pm$ SD]: 16.33 $\pm$ 22.22, early stage; 64.29 $\pm$ 145.15, intermediate stage; 121.82 $\pm$ 197.79, mature stage).
The cMUR estimates were not statistically significant considering the burst event sequences detected in the early stage, while only sporadic instances of significant presence of memory emerged in the later stages (percentage of statistically significant cMUR across all electrodes and cultures [mean $\pm$ SD]: 4.87 $\pm$ 5.49 $\%$, intermediate stage; 7.09 $\pm$ 8.51 $\%$, mature stage).
%Furthermore, these estimates appeared considerably weaker compared when the entire MUAs were considered (not shown due to brevity).
Although these findings could be affected by the limited length of burst event sequences, the fact that low statistical significance was observed also in the mature stage suggests that the bursting activity of in-vitro neural cultures exhibits little or no self-predictable patterns when analyzed in isolation.

\subsection {Heartbeat Timing Events from Healthy Subjects}

The second application regards the analysis of heartbeat timing events measured from a group of healthy subjects undergoing an experimental protocol designed to elicit different types of physiological stress \cite{javorka2018towards}. The analyzed dataset consists of the sequences of events of heartbeat timing, measured from the electrocardiogram (ECG) recordings of 61 subjects (37 female, 24 male, 17.5 $\pm$ 2.4 years old), monitored in the resting supine position during relaxation (R), in the upright body position reached after passive head-up tilt at a 45$^{\circ}$ angle (T), and in the resting position during the execution of a mental arithmetic test; further details about the protocol can be found in \cite{javorka2018towards}. The event series analyzed here are constituted by the sequences of timings of the consecutive peaks of the ECG detected at the apex of each R-wave (Fig. \ref{app_cardio}a).

%involves applying the cMUR estimator to heart rate variability data measured from healthy subjects at rest and during postural and mental stress \cite{krohova2020vascular}. Specifically, the electrocardiogram (ECG) was recorded with a sampling frequency of 1 kHz from 26 subjects underwent a protocol including a resting phase in the 15-minute supine position (R), orthostatic stress induced by passive 8-minute head-up tilt at a 45$^{\circ}$ angle (T), and mental stress induced by performing arithmetic tests 10 minutes after the tilt in the 6-minute supine position (M); 

The occurrence of each R-wave event, which marks the electrical impulse from the conduction system of the heart that represent the ventricular contraction, is modulated by neural commands stemming from the sympathetic and parasympathetic branches of the autonomic nervous system, resulting in the well-known and widely studied heart rate variability (HRV) \cite{stein1994heart}. Here we investigate how the memory capacity of heartbeat events, which is evidently associated with HRV, changes as a consequence of the transition from a resting state to conditions of postural stress and mental stress.
%evoked respectively by head-up tilt and mental arithmietics. 
To this end, we apply the cMUR estimator to sequences of $N_X =300$ events measured for each analyzed subject in the rest and stress conditions.

The results reported in Fig. \ref{app_cardio}b depict the boxplot distributions of cMUR estimated across subjects in the three  conditions, highlighting statistically significant cMUR values and differences between pairs of conditions.
A statistically significant increase in the $c\hat{\dot{M}}_{X}$ values is observed when moving from R to T stage, indicating a stronger predictive capacity of the heartbeat timings under postural stress. On the other hand, during M the cMUR values decrease significantly with respect to T, and do not exhibit statistically significant variations compared to R.
This result suggests that mental stress does not produce significant alterations in HRV predictive capacity compared to resting state.

\begin{figure} [!h]
\centering
   \includegraphics[scale=0.95]{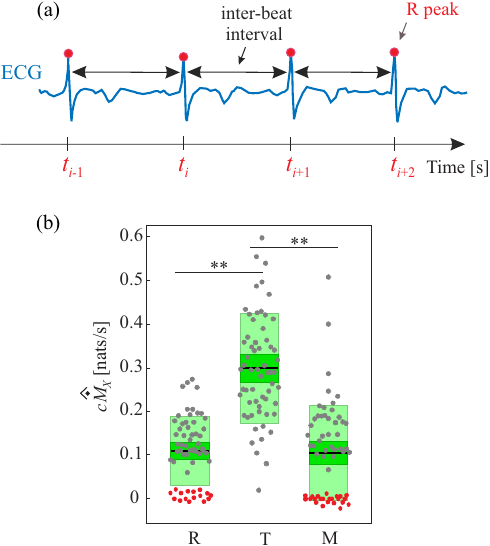}
   \caption{Application of the cMUR estimator to heartbeat timing events measured in healthy subjects during resting state (R), head-up tilt (T) and mental arithmetic (M) conditions. (a) Illustration of the extraction of the event sequences relevant to the timings of the consecutive R peaks in the ECG signal. (b) Box-plot distributions of the cMUR estimates applied over all subjects in the three conditions; significant and non-significant values detected by surrogate data analysis are denoted by gray and red dots respectively. The Wilcoxon signed-rank test was used to assess the statistical significance of the differences in the median value of the cMUR estimates between the two experimental conditions, assuming a $5\%$ significance level: $** p<0.005$. The cMUR estimator was implemented by setting: $k_{global}$ = 25, $l$ = 3, $N_U = N_X = 300$.}
    \label{app_cardio}
\end{figure}

\section{Discussion and Concluding Remarks} 

Although information-theoretic methods are widely employed in the analysis of multivariate time series within computational neuroscience and network physiology \cite{lizier2012local,wibral2014directed,porta2015wiener,faes2017information}, their application to spiking data remains relatively uncommon. This is primarily due to methodological challenges, particularly the difficulty of reliably implementing in practice measures of coupling, information transfer, and information storage. These measures are inherently designed for discrete-time processes and, when applied to neural spike trains, are usually implemented through simplified procedures (e.g., binning of the time axis) \cite{goldberg2009spike,timme2016high} that disregard the continuous-time nature of the underlying processes.
This introduces significant issues, including estimation bias, increased data requirements, and a failure to accurately capture dynamics deployed across multiple time scales (e.g., resulting in the alternance of short and long inter-spike intervals) \cite{shorten2021estimating}.
Therefore, adopting a continuous-time formalism to implement measures of information dynamics for the analysis of neural spike trains is of utmost importance. This task has been recently faced regarding the development of continuous-time information-theoretic estimators of the causal and overall coupling between spike trains, resulting in the measures of transfer entropy rate \cite{shorten2021estimating} and mutual information rate \cite{mijatovic2021information}. However, a reliable estimator of the information dynamically stored in a spike train process is still lacking, likely because the computation of information storage in continuous time is hindered by theoretical issues \cite{spinney2018characterizing}.
In this work, we fill this gap by introducing a data-efficient estimator of the memory utilization rate (MUR), the continuous-time counterpart of the measure of information storage.
The proposed estimator is specifically designed for point process data analyzed relying on the inter-event intervals, which are taken as the continuous variables that completely describe the process. The estimator is designed exploiting the popular nearest neighbor entropy estimation approach \cite{kraskov2004estimating}, modified to compute the relevant entropy and cross-entropy terms and implemented distinguishing short (single-interval) and longer history embeddings in the computation. The peculiar advantages of the method are its model-free nature, the fact that it depends only on a few non-critical parameters, and its implementation providing bias compensation and significance assessment.

The proposed MUR estimator is shown to reliably reflect the expected memory capacity in several types of simulation settings, ranging from typical Poisson processes, both memoryless and with increasingly strong history dependence, to physiologically-plausible neuronal and cardiovascular spiking processes with varying conditions of ISI interdependence. The method proved to be robust against data size and rather stable across different configurations of the estimator parameters (i.e., the embedding length $l$ and the minimum number of neighbors searched $k_{global}$); confirming previous results obtained for nearest neighbor entropy estimators \cite{shorten2021estimating, mijatovic2021information, mijatovic2022measuring}, the variance showed a tendency to decrease at increasing $k_{global}$, and the bias was efficiently compensated by the proposed correction based on subtracting the values estimated from surrogate memoryless point processes. Accordingly, we suggest working with a large number of neighbor searched as a preferable condition.
Thanks to its model-free nature, the estimator turned out to be flexible for different types of spiking dynamics: in Poisson processes with exponential ISI distribution, the MUR effectively captured both the absence of self-predictable patterns in memoryless spike trains and the presence and strength of self-predictable dynamics in trains with significant memory; in realistic models generating synchronized cortical spiking dynamics with highly heterogeneous ISIs \cite{izhikevich2003simple}, or neurally-driven heartbeat dynamics with more homogeneous ISIs showing inverse-Gaussian distribution \cite{barbieri2005point}, the MUR reliably reflected the variable predictable patterns that resulted from changes in the model parameters acting, such as the neuronal coupling strength or the rise of specific heart rate oscillations.

The application of our method to real spiking dynamics associated to central and autonomic nervous system activities allowed us to investigate the predictive memory capacity of neuronal systems explored at different scales of observation and experimental conditions.
%In the analysis of real data, the continuous-time cMUR estimator was employed to explore predictive memory capacity in experimental recordings from different sources and temporal scales. This included: (i) a microscopic examination of spiking activity in maturing in-vitro neural cultures associated with the central nervous system, and (ii) a peripheral investigation of variations in heartbeat timings known to be regulated by the autonomic nervous system. 
In the analysis of in-vitro neural cultures, we found that as neuronal cultures mature there is a concurrent increase in the memory of spiking activity at the single-electrode level captured by the cMUR estimator, which can be associated with greater synchronicity among multiple electrodes; this effect mimicked the behavior observed in our simulation of synaptically-connected neurons, corroborating the hypothesis that neural synchrony and memory are often correlated phenomena \cite{rakshit2018emergence}.
Prior studies have extensively investigated network dynamics and synchronicity in in-vitro neural cultures, primarily attributing it to the emergence of bursting activity as neural cultures mature \cite{downes2012emergence, mijatovic2021information,minati2019connectivity}. Moreover, recent research suggests that initial firing patterns in cortical in-vitro cultures can predict the developmental trajectory of network activity, indicating that self-predictable neuronal dynamics may exist from early neuronal development stages \cite{cabrera2021early}. Our findings suggest that a considerable number of electrodes in the analyzed cultures developed memory storage tendencies early on. Furthermore, these electrodes maintained their activity while increasing their memory capacities throughout subsequent developmental stages, suggesting a greater propensity for self-regularity as cultures mature. On the other side, when the cMUR estimator was applied on events representative of bursting activity, we observed a significant decrease or absence of self-predictable dynamics, suggesting that the regular patterns in the entire spiking activity are predominantly shaped by presence of bursting, even though their centers of mass do not appear in a predictable manner.

Regarding the application to heartbeat dynamics, our results indicate the presence of memory effects in the timing of cardiac electrical depolarization in all the conditions analyzed, with a marked increase of the cMUR values moving from the resting state to postural stress, and the lack of significant variations during mental stress.
These finding substantially align with a large body of literature about heart rate variability \cite{robinson1966control,stein1994heart,javorka2018towards} documenting that the variations in the interbeat intervals are under the influence of the autonomic nervous system and result in oscillations within the so-called low frequency (LF, 0.04-0.15 Hz) and high frequency (HF, 0.15-0.4 Hz depending on the respiratory rate) bands. Such variability is modulated by postural stress in a way such that the LF oscillations become predominant in the upright position, as a consequence of a shift in the sympatho-vagal balance towards sympathetic activation and parasympathetic withdrawal \cite{montano1994power}. This effect is revealed in the time domain by a decrease of the complexity of the interbeat interval series \cite{porta2006complexity,faes2019comparison}, which is here reflected by stronger memory effects (higher cMUR values).
%which demonstrated that HRV oscillatory patterns during rest are affected by cardiovascular responses to postural stressor; concretely, orthostatic stress induced by head-up-tilt is recognized to enhance sympathetic activity within the autonomic nervous system, consequently increasing sympatho-vagal balance \cite{faes2013mechanisms}. The sympathetic activity within the autonomic nervous system is primarily reflected in the low frequency (LF) band (0.04-0.15 Hz) of the spectrum, where dominant heart rate oscillations are observed during postural stress \cite{montano1994power, pernice2021comparison}. As such, it has been extensively investigated using spectral analysis or alternatively in the time domain by nonlinear model-free measures of predictability \cite{faes2019comparison}. The latter approach is often preferred because it allows detection of memory effects related to complex physiological regulation mechanisms, which may result in nonlinear correlations. Non-parametric HRV analysis frequently utilizes information-theoretic measures of information storage \cite{xiong2017entropy}, computed for discrete-time processes focusing on inter-beat intervals \cite{xiong2017entropy}. However, here we shift the paradigm by utilizing a novel continuous-time approach which revealed an increased rate of memory utilization during head-up tilt when analyzing sequences of heartbeat timings. 
On the other hand, the substantially unchanged cMUR values observed during the mental arithmetic test may be associated with the fact that the sympathetic activation evoked by mental stress is of a different type and evokes different responses than physical stress (e.g., less strong increase in cardiac output, presence of vasoconstriction and elevated blood pressure) \cite{vancheri2022mental} which are likely to alter less importantly the heartbeat interval variability.
%likely involving central commands from the upper brain centers (cortex) which independently control heartbeat and arterial compliance \cite{fauvel2000mental}. 

In conclusion, the proposed MUR estimator constitutes an important tool for the analysis of the dynamic properties of point-process data in the field of neural signal processing. This tool is complementary to recently proposed continuous-time estimators of the directed and overall coupling between spike trains like the TER \cite{shorten2021estimating} and the MIR \cite{mijatovic2021information}, and together with them provides a comprehensive means to assess the broader picture of information processing in neural spike trains, including the analysis of memory within single neural units and of signaling between pairs of units.
The integrated framework emerging from the development of these measures, which share the same computational strategy for the model-free estimation of dynamical information quantities, holds significant promise and broad applicability in the in the fields of network neuroscience and physiology where point process representations are common.
Future work should aim to extend this framework to multivariate analyses, implementing the proposed continuous-time formalism and model-free estimators for quantifying higher-order interactions within networks of neural point processes.

%We believe this approach holds significant promise and broad applicability, however especially in the fields of computational neuroscience and network physiology, where such a continuous-time formalism lends itself to efficiently quantifying information processing. Future work should aim to extend the framework by quantifying higher-order interactions among multiple point-processes, allowing for the study of generalized network structures through statistical concepts of redundancy and synergy 
%Furthermore, by quantifying the individual dynamics of a single point-process, the MUR estimator complements the univariate approach with bivariate TER and MIR point-process measures, which quantify respectively directed and undirected interactions between point-processes. This combined approach completes a robust framework for analyzing the interplay between the dynamics of individual processes and the functional connectivity emerging from pairs of processes.

%\section*{Software and Data Availability}
%The Matlab functions that perform the MUR estimation are available for free download at: \\ https://github.com/mijatovicg/MUR.

%\textcolor{red}{The dataset...?}

%\section*{References}

\bibliographystyle{IEEEtran}
\bibliography{ref} 
\end{document}